\RequirePackage{ifpdf}
\documentclass[preprint,notoc]{JHEP3}
\usepackage{epsfig}
\usepackage{slashed}
\usepackage{float}

\def\sr2{\sqrt{2}}  
 \def\to{\rightarrow} 
\def\bi{\begin{itemize}} \def\ei{\end{itemize}} 
\def\c1p{C1^\prime} \def\ta{\tilde a}  
   
  \def\ta{\tilde a} 
   
  \def\tst{\tilde t} 
 \def\tg{\tilde g}  
\def\tq{\tilde q}  \def\tw{\widetilde W} \def\tz{\widetilde
Z} 
  \def\alt{\lesssim}
\def\agt{\gtrsim} \def\be{\begin{equation}} \def\ee{\end{equation}}
\def\bea{\begin{eqnarray}} \def\eea{\end{eqnarray}}

\preprint{\vbox{OU-HEP-150831}\\\vbox{UT-15-37}}

\title{Leptogenesis scenarios for natural SUSY\\
 with mixed axion-higgsino dark matter} 
\author{Kyu Jung Bae$^{a,b}$, Howard Baer$^{a}$, Hasan Serce$^{a}$ and Yi-Fan Zhang$^{a}$\\
$^a$Dept.\ of Physics and Astronomy, University of Oklahoma, Norman, OK 73019,
USA\\
$^b$Dept.\ of Physics, University of Tokyo, Bunkyo-ku, Tokyo 113-0033, Japan\\
E-mail: \email{bae@hep-th.phys.s.u-tokyo.ac.jp}, \email{baer@nhn.ou.edu}, \email{serce@ou.edu},
\email{zyf@ou.edu} }

\abstract{Supersymmetric models with radiatively-driven electroweak 
naturalness require light higgsinos of mass $\sim 100-300$ GeV. 
Naturalness in the QCD sector is invoked via the Peccei-Quinn (PQ) axion 
leading to mixed axion-higgsino dark matter. 
The SUSY DFSZ axion model provides a solution to the SUSY $\mu$ problem and 
the Little Hierarchy $\mu\ll m_{3/2}$ may emerge as a consequence of 
a mismatch between PQ and hidden sector mass scales.
The traditional gravitino problem is now augmented by
the axino and saxion problems, 
since these latter particles can also contribute to overproduction 
of WIMPs or dark radiation, or violation of BBN constraints.
We compute regions of the $T_R$ vs. $m_{3/2}$ plane allowed by
BBN, dark matter and dark radiation constraints for various 
PQ scale choices $f_a$. 
These regions are compared to the values needed for thermal leptogenesis, 
non-thermal leptogenesis, oscillating sneutrino leptogenesis and Affleck-Dine leptogenesis. 
The latter three are allowed in wide regions of parameter
space for PQ scale $f_a\sim 10^{10}-10^{12}$ GeV which is also favored
by naturalness: $f_a\sim \sqrt{\mu M_P/\lambda_\mu }\sim 10^{10}-10^{12}$ GeV. 
These $f_a$ values correspond to axion masses
somewhat above the projected ADMX search regions.
 }

\begin{document}

\section{Introduction}
\label{sec:intro}

\subsection{Electroweak naturalness}

The recent discovery of the Higgs boson with mass $m_h\simeq 125$ GeV at LHC8~\cite{atlas_h,cms_h} 
brings with it a puzzle: why is the Higgs so light when its mass is quadratically divergent? 
Supersymmetry provides an elegant solution to this so-called naturalness
problem by providing order-by-order cancellation of quadratic divergences.
In fact, in the Minimal Supersymmetric Standard Model or MSSM, the Higgs mass is constrained
so that $m_h\alt 135$ GeV~\cite{mhiggs}: the measured mass lies comfortably below this bound.
The price to pay for a SUSY solution to the electroweak naturalness problem is that,
naively, superpartners also ought to exist at or around the weak scale $m_{\rm weak}$
as typified by the $W$, $Z$ and $h$ masses: $m_{\rm weak}\sim 100$ GeV.
However, null results from sparticle searches at LHC8 have resulted in mass limits 
within the multi-TeV regime~\cite{atlas_s,cms_s}: $m_{\tg}\agt 1.3$ TeV for $m_{\tg}\ll m_{\tq}$ and
$m_{\tg}\agt 2$ TeV for $m_{\tq}\sim m_{\tg}$. Furthermore, the somewhat large value of $m_h$ 
seems to require either well-mixed TeV scale top-squarks or 10-100 TeV top-squarks with
small mixing~\cite{h125,arbey}. These rather large sparticle mass values threaten to re-introduce 
the naturalness question: 
this time due to log divergences which emerge from the Little Hierarchy 
$m_{\rm weak}\ll m_{\rm sparticle}\sim 2-20$ TeV.
Some authors go so far as to claim the emergent Little Hierarchy leads to a crisis in 
physics~\cite{lykken}.
To proceed at a deeper level, a quantitative discussion of SUSY electroweak naturalness is warranted.

The weak scale as typified by the $Z$-boson mass is directly related to weak scale SUSY Lagrangian
parameters via the well-known scalar potential minimization condition
\bea
\frac{m_Z^2}{2} &=& \frac{(m_{H_d}^2+\Sigma_d^d)-(m_{H_u}^2+\Sigma_u^u)\tan^2\beta}{(\tan^2\beta -1)}
-\mu^2\\
&\simeq &-m_{H_u}^2-\mu^2-\Sigma_u^u(i)
\label{eq:mzs}
\eea
where $m_{H_u}^2$ and $m_{H_d}^2$ are the {\it weak scale} soft SUSY breaking Higgs masses, $\mu$
is the {\it supersymmetric} higgsino mass term and $\Sigma_u^u$ and $\Sigma_d^d$ contain
an assortment of loop corrections (labelled by index $i$) 
to the effective potential (the complete set of one-loop corrections is given in Ref.~\cite{rns}). 
The {\it electroweak fine-tuning measure} $\Delta_{\rm EW}$~\cite{rns} 
compares the largest contribution on the right-hand-side of Eq. \ref{eq:mzs} 
to the value of $m_Z^2/2$. If they are comparable, then no
unnatural fine-tunings are required to generate $m_Z=91.2$ GeV. 
The measure $\Delta_{\rm EW}$ has the advantage of being model-independent (in that 
any model yielding the same weak scale spectra will have the same value of $\Delta_{\rm EW}$).
It is also pragmatic: in computer codes that calculate the weak scale SUSY spectra, 
this is the point where actual fine-tuning occurs - usually in the form of dialing the 
required value of $\mu$ so as to ensure that $m_Z=91.2$ GeV. The implications of 
natural SUSY (SUSY spectra with low $\Delta_{\rm EW}\alt 30$~\cite{bbs}) are as follows~\cite{rns}.
\bi
\item $\mu\sim 100-300$ GeV (the lighter the better) leading to a set of light higgsinos
$\tz_{1,2}$ and $\tw_1^\pm$. The $\tz_1$ is the lightest SUSY particle (LSP) and it is
a higgsino-like WIMP which is thermally under-produced as a dark matter candidate.
\item The soft term $m_{H_u}^2$ is driven radiatively to small values $\sim -(100-300)^2$ GeV$^2$
at the weak scale (this is known as radiatively-driven naturalness or RNS).
This can always occur in models with high scale Higgs soft term non-universality.
\item The radiative corrections $\Sigma_u^u$ are $\alt (100-300)^2$ GeV$^2$. The largest of these usually 
comes from the top squark contributions $\Sigma_u^u(\tst_{1,2})$. These contributions can both
be small for well-mixed (large $A_t$) top squarks with mass $m_{\tst_1}\alt 4$ TeV and
$m_{\tst_2}\alt 10$ TeV. These same conditions lift $m_h$ into the 125 GeV vicinity.
\ei

While a value of low $\Delta_{\rm EW}$ seems like a necessary condition for SUSY naturalness, 
the question is: is it also sufficient? Does it embody high scale fine-tuning as well? 
The answer given in Ref's~\cite{comp,seige} is that, yes, Eq. \ref{eq:mzs} provides a complete 
portrayal of SUSY electroweak fine-tuning. 
\bi
\item An often-invoked alternative known as Higgs mass 
fine-tuning requires no large cancellations in contributions to the Higgs boson
mass: $m_h^2\sim \mu^2+m_{H_u}^2+\delta m_{H_u}^2$ and thus requires $\delta m_{H_u}^2\alt m_h^2$. 
The value of $\delta m_{H_u}^2$ can be calculated by integrating the renormalization 
group equation for the soft term $m_{H_u}^2$. A back-of-the-envelope evaluation
leads to $\delta m_{H_u}^2\sim -\frac{3f_t^2}{8\pi^2}(m_{Q_3}^2+m_{U_3}^2+A_t^2)\ln\left(\Lambda^2/m_{\rm SUSY}^2 \right)$ where $\Lambda$ is the high scale usually 
taken to be $m_{\rm GUT}$ and $m_{\rm SUSY}\sim 1$ TeV. 
However, this overly simplified expression neglects the fact that $m_{H_u}^2$ itself 
feeds into the evaluation of $\delta m_{H_u}^2$. In fact, the larger the value of $m_{H_u}^2(\Lambda )$, then the larger is the cancelling correction $\delta m_{H_u}^2$. This evaluation violates the 
{\it fine-tuning rule}~\cite{seige}: to avoid over-estimates of fine-tuning, 
first combine {\it dependent} contributions to any observable quantity. 
By following the fine-tuning rule, then instead the two terms on the RHS of 
$m_h^2=\mu^2+\left( m_{H_u}^2(\Lambda )+\delta m_{H_u}^2 \right)$ should be comparable
to $m_h^2$. Since $m_{H_u}^2(\Lambda )+\delta m_{H_u}^2=m_{H_u}^2({\rm weak})$, then the 
Higgs mass fine-tuning conditions lead to the same as those for low $\Delta_{\rm EW}$.  
\item EENZ/BG fine-tuning~\cite{eenz,bg} $\Delta_{\rm BG}=\max_i\left|\frac{\partial\ln m_Z^2}{\partial\ln p_i}\right|$
measures the sensitivity of $m_Z^2$ to high scale parameters $p_i$. The usual application of 
$\Delta_{\rm BG}$ is to multi-parameter effective theories where the various $p_i$ parametrize
our ignorance of the nature of the hidden sector which serves as an arena for SUSY breaking.
By recognizing that  in any SUGRA theory the soft terms are all calculated as multiples of
the gravitino mass $m_{3/2}$, then the $Z$ mass can be expressed in terms of high scale parameters as 
$m_Z^2\simeq -2 \mu^2(\Lambda )+a\cdot m_{3/2}^2$. 
Since $\mu$ hardly evolves, then $a\cdot m_{3/2}^2\simeq m_{H_u}^2({\rm weak})$ so that a low
value of $\Delta_{\rm BG}$ leads again to the same requirements as a low value of $\Delta_{\rm EW}$.
\ei

\subsection{Naturalness in the QCD sector}

In QCD, in the limit of two light quarks $u,\ d$, one has an approximate global $U(2)_L\times U(2)_R$ chiral symmetry which can be
recast as $U(2)_V\times U(2)_A$. The vector symmetry leads to well-known SU(2) of isospin along with baryon number conservation.
The axial $U(2)_A$ symmetry is spontaneously broken and naively leads to four instead of three light pions as pseudo-Goldstone bosons.
Weinberg suggested~\cite{u1a} the U(1)$_A$ symmetry was somehow not respected and indeed this viewpoint was vindicated by 't~Hooft's
discovery of the QCD $\theta$ vacuum. A consequence of this resolution of the U(1)$_A$ problem is that the QCD Lagrangian should
contain a $C$ and $CP$-violating term
\be
{\cal L}_{\rm QCD}\ni \frac{\bar{\theta}}{32\pi^2}G_{\mu\nu}^A\tilde{G}^{A\mu\nu}
\ee
(where $\bar{\theta}=\theta+\arg\ (\det{\cal M})$ and where $G_{\mu\nu}^A$ is the gluon field strength tensor). 
Measurements of the neutron EDM require $\bar{\theta}\alt 10^{-10}$ which leads to the QCD naturalness problem (also known as
the strong $CP$ problem): why is this term - which should be present - so tiny? Peccei and Quinn~\cite{pq} introduced
an additional global PQ symmetry which is spontaneously broken at scale $v_{PQ}\sim 10^{10}$ GeV leading to 
a quasi-visible~\cite{ksvz,dfsz} axion~\cite{ww}. Introduction of the axion field allows the offending
$CP$-violating term to dynamically settle to zero, thus saving the day for QCD. The required axion field ought to 
have a mass $m_a\sim 620$ $\mu$eV$\left(\frac{10^{10}\ {\rm GeV}}{f_a/N}\right)$ 
where $N$ is the color anomaly of the PQ symmetry ($N=1$ for KSVZ~\cite{ksvz} and $N=6$ for DFSZ~\cite{dfsz}).

The axion can be produced via axion field coherent oscillations in the early universe and serves as a candidate
for cold dark matter~\cite{axdm}. In a SUSY context, the axion should be accompanied by the spin-1/2 $R$-parity-odd axino $\ta$ and the
spin-0 $R$-even saxion field $s$. In supergravity models, the soft breaking saxion mass $m_s\sim m_{3/2}$. 
The axino mass is more model dependent~\cite{cl,kim} but is usually also expected to be $m_{\ta}\sim m_{3/2}$.
The axion, saxion and axino interactions are all suppressed by the inverse of the axion decay constant
$f_a$ where $f_a\sim v_{PQ}$.

\subsection{Naturalness and the $\mu$-problem}

While the axion plays a crucial role in solving the QCD naturalness problem, it also plays a role in the electroweak
naturalness problem. While the soft term $m_{H_u}^2$ can be radiatively driven to small negative values by the large
top-quark Yukawa coupling, the $\mu$ parameter in Eq. \ref{eq:mzs} also needs to be tamed. Since it is supersymmetric
and not SUSY breaking, naively one expects $\mu\sim m_{\rm GUT}$ or $M_{P}$ (the reduced Planck mass). In contrast, naturalness
requires $\mu\sim m_{\rm weak}$. The SUSY DFSZ axion provides an elegant resolution of this so-called SUSY $\mu$ problem~\cite{kn}
in that the Higgs superfields now carry PQ charge which forbids the appearance of the $\mu$ term in the superpotential.
Upon spontaneous breaking of the PQ symmetry, then in the SUSY DFSZ axion model the $\mu$ term is regenerated with a value
$\mu\sim \lambda_\mu f_a^2/M_P$. This is in contrast to the SUSY soft terms which gain a mass $m_{soft}\sim m_{3/2}\sim m_{\rm hidden}^2/M_P$ where $m_{\rm hidden}$ is the hidden sector mass scale.
In such a case, the emerging Little Hierarchy $\mu\ll m_{3/2}$ is just a consequence of a mismatch between 
hidden sector and PQ sector intermediate mass scales $f_a\ll m_{\rm hidden}$. 
In fact, in models such as the MSY SUSY axion model~\cite{msy}, 
the PQ symmetry is broken radiatively as a consequence of SUSY breaking leading naturally to $\mu\sim 100-300$ GeV 
whilst $m_{3/2}\sim 2-10$ TeV~\cite{Bae:2014yta}. 
As a by-product of radiative PQ breaking, intermediate scale Majorana masses are also
induced $m_N\sim f_a$ leading to see-saw neutrinos~\cite{seesaw}. In this scenario, then, the PQ breaking scale $f_a$ 
plays a role in determining the axion, the higgsino and the Higgs masses!

\subsection{Dark matter in SUSY with electroweak and QCD naturalness}

In a highly natural model where the electroweak sector is stabilized by SUSY, the QCD sector is stabilized by the axion, 
the $\mu$ problem is resolved by PQ-charged Higgs fields and the Little Hierarchy $\mu\ll m_{3/2}$ emerges from radiative PQ
breaking, then one expects dark matter to be composed of an axion-higgsino admixture: {\it i.e.} two dark matter particles.
The favored axion scale $f_a\sim \sqrt{\mu M_P/\lambda_\mu }\sim 10^{10-12}$ GeV (for $\lambda_\mu\sim 0.01-1$) 
which is somewhat below the range of $f_a$ currently being explored by the axion dark matter search experiment, ADMX~\cite{admx}.
Regarding the higgsino-like WIMPs, their relic abundance calculation is seriously modified from the usual thermal
production picture~\cite{kimrev}. To be sure, the WIMPs are still produced thermally, but they can also arise via axino and saxion
production in the early universe, followed by cascade decays which terminate in WIMPs. 
While axinos can be produced thermally, saxions can be produced both thermally and via coherent oscillations.
If too many WIMPs are produced via heavy particle decays, then they may undergo a re-annihilation process~\cite{ckls,blrs}. 
Furthermore, axions can also be produced thermally or via saxion decays. 
The latter leads to injection of dark radiation parametrized by the effective number of additional
neutrinos present in the cosmic soup: $\Delta N_{\rm eff}$. 
Current bounds from the Planck experiment require $N_{\rm eff}=3.15\pm 0.23$~\cite{planck}. 
(To be conservative, here we require merely $\Delta N_{\rm eff}\alt 1$ in our results.)
With an assortment of interwoven production and decay processes occurring, an accurate estimate of the ultimate
mixed axion-higgsino dark matter content requires simultaneous solution of eight coupled Boltzmann equations which track 
the abundance of radiation, WIMPs, thermal- and oscillation-produced axions, thermal- and oscillation-produced saxions, axinos
and gravitinos~\cite{dfsz2}. 

Results vary radically depending on whether one is in a hadronic (KSVZ)~\cite{ksvz} SUSY axion model
or DFSZ~\cite{dfsz} SUSY axion model. 
In the former KSVZ case, thermal production of axinos and saxions is proportional to 
the re-heat temperature $T_R$~\cite{axnoprod_k} while decay modes arise from heavy quark induced loop 
diagrams due to the superpotential term
\be
W_{\rm KSVZ}\ni m_Qe^{A/f_a}QQ^c
\ee 
where $Q$ stands for intermediate-scale heavy quark superfields with $m_Q\sim f_a$.
In SUSY DFSZ, axino and saxion thermal productions are different from those in SUSY KSVZ since the 
axion superfield has tree level couplings which are proportional to the SUSY $\mu$ parameter~\cite{axnoprod_d1,axnoprod_d2,axnoprod_d3}:
\be
W_{\rm DFSZ}\ni \mu e^{-2A/f_a}H_u H_d .
\label{eq:superptl}
\ee
Due to this interaction, thermal production of axions, axinos and saxions is largely independent of $T_R$ unless $T_R\lesssim \mu$~\cite{axnoprod_d2}.
Decays also dominantly proceed through this tree level coupling so the axino and saxion tend to be shorter-lived than in the KSVZ case.

\subsection{Connection to baryogenesis}

One of the major mysteries of particle physics and cosmology is the origin of the matter-anti-matter asymmetry
as embodied by the measurement of the baryon-to-photon ratio~\cite{pdb}
\be
\eta_B\equiv \frac{n_B}{n_\gamma}\simeq (6.2\pm 0.5)\times 10^{-10}\ \ \ \ (95\%\ CL) .
\ee
$\eta_B$ is determined both from light element production in Big Bang Nucleosynthesis (BBN) 
and also from CMB measurements. Alternatively, this is sometimes expressed as the baryon-to-entropy ratio
\be
\frac{n_B}{s}\simeq 10^{-10}
\ee
where $s\simeq 7.04n_\gamma$ in the present epoch.

Production of the baryon asymmetry of the Universe or BAU requires mechanisms which satisfy the three Sakharov criteria:
1. baryon number violation, 2. $C$ and $CP$ violation and 3. a departure from thermal equilibrium. Early proposals
such as Planck scale or GUT scale baryogenesis seem no longer viable since the BAU would have been inflated
away during the inflationary epoch of the Universe. Alternatively, most modern proposals for developing the BAU
take place after the end of the inflationary epoch, at or after the era of re-heating which occurs 
around the re-heat temperature $T_R$. In fact, the SM contains all the ingredients for successful
{\it electroweak} baryogenesis since baryon (and lepton) number violating processes can take place at large rates
at high temperature $T>T_{\rm weak}\sim 100$ GeV via sphaleron processes~\cite{krs}. 
Unfortunately, these first order phase transition
effects require a Higgs mass $\alt 50$ GeV, and so has been excluded for many years. By invoking supersymmetry, then new
possibilities emerge for electroweak baryogenesis. However, successful SUSY electroweak baryogenesis seems to require
a Higgs mass $m_h\alt 113$ GeV and a right-handed top-squark $m_{\tst_R}\alt 115$ GeV~\cite{cw}. 
These limits can be relaxed 
to higher values so long as other sparticle/Higgs masses such as $m_A\gg 10$ TeV. Such heavy Higgs masses are not allowed
if we stay true to our guidance from naturalness: after all, Eq. \ref{eq:mzs} requires $m_{H_d}^2/\tan^2\beta\alt m_Z^2/2$.
For heavy Higgs masses, then $m_A\sim m_{H_d}$ and then from naturalness we find $m_A\alt 4-8$ TeV 
(depending on $\tan\beta$)~\cite{savoy1}.

In Sec. \ref{sec:survey}, we survey several leptogenesis mechanisms as the most promising baryogenesis mechanisms: 
1. thermal leptogenesis~\cite{thlepto,bdp}, 
2. non-thermal leptogenesis via inflaton decay~\cite{ntlepto} 
3. leptogenesis from oscillating sneutrino decay~\cite{ADlepto,Hamaguchi:2001gw} and
4. leptogenesis via an Affleck-Dine condensate~\cite{AD,drt,ADlepto}.

Each of these processes requires some range of re-heat temperature $T_R$ and gravitino mass $m_{3/2}$, and indeed
some of them run into conflict with the so-called cosmological gravitino problem~\cite{gravprob}.
In this case, gravitinos can be thermally produced in the early universe at a rate proportional to $T_R$~\cite{gravprod}. 
If $T_R$ is too high then too much dark matter arises from thermal gravitino production followed by
cascade decays to the LSP. Also, even if dark matter abundance constraints are respected, if the gravitino is too
long-lived, then it may decay after the onset of BBN thus destroying the successful BBN predictions of the light element abundances~\cite{olive,moroi,jedamzik}.

In the case of natural SUSY with mixed axion-higgsino dark matter, then similar constraints 
arise from axino and saxion production:
WIMPs or axions can be overproduced, or light element abundances can be destroyed by late decaying axinos and saxions.
After a brief review of the several leptogenesis mechanisms in Sec. \ref{sec:survey}, 
in Sec. \ref{sec:tr_m32} we show constraints on leptogenesis in the $T_R$ vs. $m_{3/2}$ planes 
assuming a natural SUSY spectrum.
In Sec. \ref{sec:tr_fa}, we show corresponding results in the $T_R$ vs. $f_{a}$ planes.
We vary the PQ scale $f_a$ from values favored by naturalness $f_a\sim 10^{10}-10^{12}$ GeV to much higher values.
While thermal leptogensis mechanism is quite constrained depending on $m_{3/2}$ and $T_R$, the 
latter three mechanisms appear plausible over a wide range of $T_R$, $m_{3/2}$ and $f_a$ values which are consistent with naturalness.
A summary and some conclusions are presented in Sec. \ref{sec:conclude}.

\section{Survey of some baryogenesis mechanisms}
\label{sec:survey}

\subsection{Thermal leptogenesis (THL)}

Thermal leptogenesis~\cite{thlepto,bdp,THLreviews} is a baryogenesis mechanism which relies on the introduction of 
three intermediate mass scale right hand singlet neutrinos $N_i$ ($i=1-3$) so that the 
(type I) see-saw mechanism~\cite{seesaw} elegantly generates a very light spectrum of usual neutrino masses. 
The superpotential is given by
\be
W\ni \frac{1}{2}M_iN_iN_i+h_{i\alpha}N_iL_\alpha H_u
\ee
where we assume a basis for the $N_i$ masses which is diagonal and real. The index $\alpha$ denotes the lepton doublet 
generations and $h_{i\alpha}$ are the neutrino Yukawa couplings.
From the see-saw mechanism, one expects a spectrum of three sub-eV mass Majorana neutrinos $m_1$, $m_2$ and $m_3$ 
and three heavy neutrinos $M_1<M_2<M_3$ where in GUT-type theories one typically expects $M_3\sim 10^{15}$ GeV. 
If the heavy neutrino masses are hierarchical (as assumed here) like the quark masses, 
then one might expect $M_1/M_3\sim m_u/m_t\sim 10^{-5}$ and so perhaps $M_1\sim 10^{10}$ GeV. 

After inflation, then it is assumed the Universe re-heats to a temperature $T_R\agt M_1$
thus creating a thermal population of $N_1$s. The $N_1$ decay asymmetrically as 
$N_1\to LH_u$ vs. $\bar{L} \bar{H}_u$ due to interference between 
tree and loop-level decay diagrams which include CP violating interactions.
The CP asymmetry factor 
\be
\epsilon_1 \equiv \frac{\Gamma (N_1\to LH_u)-\Gamma (N_1\to \bar{L}\bar{H}_u)}{\Gamma_{N_1}}
\ee
is calculated to be~\cite{crv}
\be
\epsilon_1\simeq \frac{3}{8\pi}\frac{M_1}{\langle H_u\rangle^2} m_{\nu_3}\delta_{\rm eff}
\label{eq:eps1}
\ee
where $\langle H_u\rangle \simeq 174\ {\rm GeV}\sin\beta$ and $\delta_{\rm eff}$ is an effective 
CP-violating phase which depends on the MNS matrix elements and which is expected to be $\delta_{\rm eff}\sim 1$.
For hierarchical heavy neutrinos, one expects 
\be
\epsilon_1\sim 2\times 10^{-10}\left(\frac{M_1}{10^6\ {\rm GeV}}\right)\left(\frac{m_{\nu_3}}{0.05\ {\rm eV}}\right)\delta_{\rm eff} .
\ee

The ultimate lepton asymmetry requires evaluation via a coupled 
Boltzmann equation calculation~\cite{plumacher}. 
The lepton-number-density to entropy ratio is given by
\be
\frac{n_L}{s}=\kappa\epsilon_1\frac{n_{N_1}}{s}=\kappa\frac{\epsilon_1}{240}
\ee
where the co-efficient $\kappa$ accounts for washout effects and the efficiency of thermal $N_1$ production.
Numerical evaluations of $\kappa$ imply $\kappa\simeq 0.05-0.3$.

The induced lepton asymmetry becomes converted to a baryon asymmetry via $B$- and $L$- violating but 
$B-L$ conserving sphaleron interactions~\cite{krs,ks_ht}.
The ultimate baryon asymmetry is given by~\cite{buchmuller}
\be
\frac{n_B}{s} \simeq 0.35\frac{n_L}{s}\simeq 0.3\times 10^{-10}\left(\frac{\kappa}{0.1}\right)
\left(\frac{M_1}{10^9\ {\rm GeV}}\right) \left(\frac{m_{\nu_3}}{0.05\ {\rm eV}}\right)\delta_{\rm eff}
\ee
provided that $T_R$ is large enough that the $N_1$ are efficiently produced by thermal interactions: 
$T_R\agt M_1$. 
Naively, this requires $T_R\agt 10^{10}$ GeV although detailed calculations~\cite{plumacher} 
allow for $T_R\agt 1.5\times 10^9$ GeV.
This rather large lower bound on $T_R$ potentially leads to conflict with the gravitino problem and 
violation of BBN bounds or overproduction of dark matter.
In the event that late-decaying relics inject entropy after $N_1$ decay is complete, then $n_L/s$ is modified by
an entropy-dilution factor $r$: $n_L/s\to n_L/rs$.

\subsection{Non-thermal leptogenesis via inflaton decay (NTHL)}

As an alternative to thermal leptogenesis, 
non-thermal leptogenesis posits a large branching fraction of the inflaton field $\chi$ into $N_1N_1$:
$\chi\to N_1N_1$ which is followed by asymmetric $N_1$ decay to (anti-)leptons as before.
In this case, the $N_1$ number-density-to-entropy-density ratio is given by~\cite{ntlepto,hamaguchi}
\bea
\frac{n_{N_1}}{s}&\simeq &\frac{\rho_{\rm rad}}{s}\frac{n_\chi}{\rho_\chi}\frac{n_{N_1}}{n_\chi}\\
&=&\frac{3}{4}T_R\times\frac{1}{m_\chi}\times2 B_r =\frac{3}{2}B_r\frac{T_R}{m_\chi}
\eea
where $\rho_{\rm rad}$ is the radiation density once reheating has completed and 
$\rho_\chi$ is the energy density stored
in the inflaton field just before inflaton decay. Thus, $\rho_{\rm rad}\simeq \rho_\chi$ and 
$\rho_\chi\simeq m_\chi n_\chi$.
Here also $B_r$ is the inflaton branching fraction into $N_1N_1$.
The lepton-number-to-entropy ratio is then given by $n_L/s\simeq \epsilon_1n_{N_1}/2$ where $\epsilon_1$ is as in Eq. \ref{eq:eps1}.

The lepton number asymmetry is converted to a baryon asymmetry via sphaleron interactions as before:
\be
\frac{n_B}{s}\simeq 0.35\frac{n_L}{s}
\ee
so that finally
\be
\frac{n_B}{s}\simeq 0.5\times 10^{-10} B_r\left(\frac{T_R}{10^6\ {\rm GeV}}\right)
\left(\frac{2M_1}{m_\chi}\right)\left(\frac{m_{\nu_3}}{0.05\ {\rm eV}}\right)\delta_{\rm eff} 
\ee
where $\delta_{\rm eff}$ is the same phase as given above.
The resultant baryon asymmetry can match data provided $m_\chi>2M_1$ and that the branching fraction is nearly maximal. 
Under these conditions, a re-heat temperature $T_R\agt 10^6$ GeV is required. For $T_R\alt 10^6$ GeV, then
$\rho_{\rm rad}$ and consequently $\rho_\chi$ are reduced so that there is insufficient energy stored in the inflaton field to
generate the required $n_{N_1}$ number density.

\subsection{Leptogensis from oscillating sneutrino decay (OSL)}
\label{ssec:oscsn}

In the previous two mechanisms, right-handed neutrinos and sneutrinos are produced by thermal scattering or inflaton decay.
On the other hand, for right sneutrinos\footnote{The spin-0 partners of right-hand neutrinos.}, 
coherent oscillation can be a dominant production process.
The decay of oscillating sneutrino produces lepton asymmetry which is given by~\cite{Hamaguchi:2001gw}
\be
n_L=\epsilon_1 M_1\left| \widetilde{N}_{1d}\right|^2,
\ee
where $\widetilde{N}_{1d}$ is the sneutrino amplitude when it decays.
The CP asymmetry factor $\epsilon_1$ is the same as thermal leptogenesis which is shown in Eq.~\ref{eq:eps1}.

Once the universe is dominated by sneutrino oscillation, pre-existing relics are mostly diluted away and the universe is reheated again by sneutrino decay at $H=\Gamma_{N_1}$, where $\Gamma_{N_1}$ is the sneutrino decay rate.
The decay temperature $T_{N_1}$ is determined by 
\be
T_{N_1}=\left(\frac{90}{\pi^2 g_*}\right)^{1/4}\sqrt{M_P\Gamma_{N_1}},
\ee
while the entropy density is given by
\be
s=\frac{2\pi^2}{45}g_*T_{N_1}^3,
\ee
where $g_*$ is the number of degree of freedom at $T=T_{N_1}$.
From these relations, one finds the lepton-number-to-entropy ratio:
\begin{eqnarray}
\frac{n_L}{s}&=&\frac{3}{4}\epsilon_1\frac{T_{N_1}}{M_1}\nonumber\\
&\simeq&1.5\times 10^{-10}\left(\frac{T_{N_1}}{10^6\mbox{ GeV}}\right)
\left(\frac{m_{\nu3}}{0.05\mbox{ eV}}\right)\delta_{\rm eff}.
\end{eqnarray}
The baryon asymmetry is obtained via sphaleron process, and thus baryon number is given by
\be
\frac{n_B}{s}\simeq0.35\frac{n_L}{s}.
\ee
Thus, enough baryon number can be generated for $T_{N_1}\gtrsim 10^{6}$ GeV.

In this scenario, it is interesting that the effective reheat temperature is ${\cal O}(T_{N_1})$ for thermal relic particles, since sneutrino domination dilutes pre-existing particles when it decays~\cite{Hamaguchi:2001gw}.\footnote{It is assumed that inflaton decay after sneutrino oscillation starts. If sneutrino oscillation starts after inflaton decay, effective reheat temperature is given by $2T_{N_1}(T_R/T_{R_C})$ where $T_{R_C}$ is the temperature at which sneutrino oscillation starts.}
Therefore, in the numerical analyses of Sec's.~\ref{sec:tr_m32} and  \ref{sec:tr_fa}, 
we consider $T_{N_1}$ a reheat temperature for production of gravitinos, axinos and saxions in the case of leptogenesis from oscillating sneutrino decay.

\subsection{Affleck-Dine leptogenesis (ADL)}
\label{ssec:AD}

The last mechanism for baryogenesis is known as Affleck-Dine (AD)~\cite{AD,drt} 
leptogenesis. 
AD leptogenesis makes use of the $LH_u$ flat direction in the scalar potential~\cite{my,drt,martin}.
This direction is lucrative in that it is not plagued by $Q$-balls which are
problematic for flat directions carrying baryon number~\cite{em} and also because
the rate for baryogenesis can be linked to the mass of the lightest neutrino, leading to
a possible consistency check via observations of neutrinoless double beta decay ($0\nu\beta\beta$)~\cite{beta}.

In the case of the $LH_u$ flat direction, $F$-flatness is only broken by higher dimensional
operators which also give rise to neutrino mass via the see-saw mechanism~\cite{seesaw}:
\be
W\ni \frac{1}{2M_i}(L_i H_u)(L_i H_u)
\label{eq:W_F}
\ee
where $M_i$ is the heavy neutrino mass scale. The most efficient direction is that for which
$i=1$ corresponding to the lightest neutrino mass: $m_{\nu_1}\sim\langle H_u\rangle^2/M_1$
in a basis where the neutrino mass matrix is diagonal.
The Affleck-Dine field $\phi$ then occurs as
\be
\tilde{L}_1=\frac{1}{\sqrt{2}}\left(\begin{array}{c} \phi\\ 0\end{array}\right)\ \ \ 
H_u=\frac{1}{\sqrt{2}}\left(\begin{array}{c} 0 \\ \phi \end{array}\right) .
\ee

The scalar potential is given by
\be
V=V_{SB}+V_H+V_{TH}+V_F
\ee
where
\bea
V_{SB}&=&m_\phi^2|\phi |^2 +\frac{m_{\rm SUSY}}{8M}(a_m\phi^4 +h.c.)\\
V_H&=&-c_H H^2|\phi|^2+\frac{H}{8M}(a_H\phi^4 +h.c. )\\
V_{TH}&=&\sum_{f_k|\phi|<T}c_kf_k^2T^2|\phi|^2+ \frac{9\alpha_s^2(T)}{8}T^4\ln\left(
\frac{|\phi|^2}{T^2}\right)\ \ \ {\rm and} \\
V_F&=&\frac{1}{4M^2}|\phi|^6 .
\eea
The first contribution $V_{SB}$ is the SUSY breaking contribution where 
$m_\phi^2= (\mu^2 +m_{H_u}^2+m_L^2)/2$~\cite{allahverdi}. In natural SUSY, we
expect $|\mu |\sim |m_{H_u}| \sim m_Z$ in contrast to  $m_L\sim m_{\rm SUSY}\sim 2-10$ TeV
in accord with LHC8 limits.\footnote{In gravity mediation, it is natural to have $m_{\rm SUSY}\sim m_{3/2}$. In our benchmark study in Sec.~\ref{sec:tr_m32}, however, $m_{\rm SUSY}$ for SUSY particle spectrum is fixed while physical gravitino mass $m_{3/2}$ varies from 1 TeV to 100 TeV.}
The second contribution arises from SUSY breaking during inflation~\cite{drt} 
where $3H_{I}^2m_{\rm GUT}^2\simeq |F_\chi|^2$ with $H_I$ being the Hubble constant during inflation and where 
$F_\chi$ is the inflaton $F$-term which fuels inflation and $\chi$ is the inflaton field.
In the expression $V_H$, the coefficient $c_H$ may be $>0$ for a non-flat 
Kahler metric (which is to be expected in general).
This term provides an instability of the potential at $|\phi|=0$ and for $c_H>0$, then 
a large VEV of $\phi$ can form with value $\langle\phi\rangle\sim \sqrt{MH_I}$
where $H_I\gg m_\phi$ and where $\arg(\phi )=[(-\arg(a_H)+(2n+1)\pi]/4$ for $n=0-3$.
The second term in $V_H$ is 
the Hubble-induced trilinear SUSY breaking term.
The term $V_F$ is the up-lifting $F$-term contribution arising from the 
higher-dimensional operator \ref{eq:W_F}. Lastly, the term $V_{TH}$ arises from thermal
effects~\cite{aco,ADth}.
The first term is generated when the light particle species which couple to the AD field are produced in the 
thermal plasma, while the second term is generated by effective gauge coupling running from heavy effective mass of particles which couple to the AD field.
Here, $f_k$ represents the Yukawa/gauge
couplings of $\phi$ and $c_k$ is expected $\sim 1$.

The equation of motion for the AD field is given by
\be
\ddot{\phi}+3H\dot{\phi}+\frac{\partial V}{\partial\phi^*}=0 
\ee
which is the usual equation for a damped harmonic oscillator.
Once the AD condensate forms, then the universe continues expansion and the Hubble-induced 
terms decrease. The minimum of the potential decreases as does the value of the
condensate. When $H$ decreases to a value~\cite{fhy}
\be
H_{\rm osc}=\max\left[ m_\phi,H_i,\alpha_2T_R\left(\frac{9M_P}{8M}\right)^{1/2}\right]
\label{eq:Hosc}
\ee
(where $H_i=min\left[\frac{1}{f_i^4}\frac{M_PT_R^2}{M^2},(c_i^2f_i^4M_PT_R^2)^{1/3}\right]$)
then the AD field begins to oscillate, and a non-zero lepton number arises: 
$n_L=\frac{i}{2}(\dot{\phi}^*\phi-\phi^*\dot{\phi})$ governed by
\be
\dot{n}_L+3Hn_L=\frac{m_{\rm SUSY}}{2M}Im(a_m\phi^4)+\frac{H}{2M}Im(a_H\phi^4) .
\label{eq:nLdot}
\ee

The first term on the RHS of Eq. \ref{eq:nLdot} is dominant and using 
$d/dt(R^3n_L)=R^3\dot{n}_L+3R^3Hn_L$ we can integrate from early times up to $t=1/H_{\rm osc}$
to find
\be
n_L=\frac{m_{\rm SUSY}}{2M}Im(a_m\phi^4)t_{\rm osc}
\ee
where $t_{\rm osc}=2/3H_{\rm osc}$ for an oscillating field/matter-dominated universe.
The lepton-number-to-entropy ratio is assumed conserved once the era of re-heat is completed:
\be
\frac{n_L}{s}=\frac{M T_R}{12M_P^2}\left(\frac{m_{\rm SUSY}}{H_{\rm osc}}\right)\delta_{\rm eff} .
\ee 
This quantity has the virtue of being $T_R$ independent if $H_{\rm osc}$ is determined by the third
(thermal) contribution in Eq. \ref{eq:Hosc}~\cite{fhy}. The lepton asymmetry is then converted
to a baryon asymmetry via sphaleron interactions
\be
\frac{n_B}{s}\simeq 0.35\frac{n_L}{s} .
\ee
Replacing $M$ by $\langle H_u\rangle^2/m_{\nu_1}$, then it is found~\cite{fhy} that a
baryon-to-entropy ratio $n_B/s\sim 10^{-10}$ can be developed roughly independent of 
$T_R$ for $T_R\agt 10^5$ GeV for $m_{\nu_1}\sim 10^{-8}$ eV and for $m_{\rm SUSY}\sim 10$ TeV.

\section{Constraints in the $T_R$ vs. $m_{3/2}$ plane for various $f_a$}
\label{sec:tr_m32}

To compute the mixed axion-WIMP dark matter abundance in SUSY axion models, we
adopt the eight-coupled Boltzmann equation computation of Ref's~\cite{bls,bbl,dfsz2}.
In that treatment, one begins at temperature $T=T_R$ and tracks the energy densities of
radiation, WIMPs, gravitinos, axinos, saxions (CO- and thermally-produced) 
and axions (CO-, thermally- and saxion decay-produced). Whereas WIMPs quickly reach thermal equilibrium
at $T=T_R$, the axinos, saxions, axions and gravitinos do not, even though they 
are still produced thermally. 
In SUSY KSVZ, the axino, axion and saxion thermal production rates are all proportional to
$T_R$~\cite{axnoprod_k} while in SUSY DFSZ model they are largely independent of $T_R$~\cite{axnoprod_d2}.
The calculation depends sensitively on the sparticle mass spectrum, 
on the re-heat temperature $T_R$, on the gravitino mass $m_{3/2}$ and on the PQ model 
(KSVZ or DFSZ), the PQ parameters $f_a$, the axion mis-alignment angle $\theta_i$, the saxion
angle $\theta_s$ (where the initial saxion field value is given as $s=\theta_s f_a$) 
and on a parameter $\xi_s$ which accounts for the model-dependent saxion-to-axion coupling~\cite{cl}.
Here, we adopt the choices $\xi_s =0$ ($s\to aa,\ \ta\ta$ decays turned off) or $\xi_s =1$
($s\to aa,\ \ta\ta$ decays turned on). 

The calculation depends sensitively on the axino, saxion
and gravitino decay rates. The gravitino decay rates are adopted from Ref.~\cite{gravdecay} while the
axino and saxion decay rates are given in Ref's~\cite{blrs,AY} for 
SUSY KSVZ and in Ref.~\cite{dfsz1} for SUSY DFSZ. The axino decays via loops involving the heavy quark $Q$
field such that $\ta\to g\tg$, $\tz_i\gamma$ and $\tz_i Z$ in SUSY KSVZ. 
In SUSY DFSZ, the axino
couples directly to Higgs superfields yielding faster decay rates into 
gauge/Higgs boson plus gaugino/higgsino states. In SUSY KSVZ, the saxion decays via $s\to gg$, $\tg\tg$ and,
when $\xi_s =1$, also to $aa$ and $\ta\ta$ (if kinematically allowed).
The decay $s\to aa$ leads to production of dark radiation as parametrized by $\Delta N_{\rm eff}$.
In SUSY DFSZ, the saxion decays directly to gauge- or Higgs-boson pairs or to 
gaugino/higgsino pairs~\cite{dfsz1}. 
If $\xi_s =1$, then also $s\to aa$ or $\ta\ta$.
In the case where axinos or saxions decay to SUSY particles (leading to WIMPs), 
then WIMPs may re-annihilate.

For the SUSY mass spectrum, we generate a 
natural SUSY model within the context of the 2-extra parameter non-universal Higgs (NUHM2)
model with $m_0=5$ TeV, $m_{1/2}=0.7$ TeV, $A_0=-8.4$ TeV and $\tan\beta =10$. 
We take $\mu =125$ GeV and $m_A=1$ TeV. 
The spectrum is generated using IsaSUGRA 7.84~\cite{isasugra}. 
The value of $m_{\tg}=1.8$ TeV so the model
is safely beyond LHC8 constraints. The value of $m_h=125$ GeV and $\Delta_{\rm EW}=20$ so the model is
highly natural. Higgsino-like WIMPs with mass $m_{\tz_1}=115.5$ GeV are thermally underproduced so that 
$\Omega_{\tz_1}^{\rm TP}h^2=0.007$ using IsaReD~\cite{isared}. 
In all frames, we take $m_{\ta}= m_s=m_{3/2}$ as is roughly expected in gravity-mediated 
SUSY breaking models~\cite{cl,kim}.
Since we take $m_{\ta}=m_s$, then $s\to\ta\ta$ decays are never a factor in our results.

In all plots, the light-blue region corresponds to the parameter space where all BBN, DM and dark radiation 
constraints are satisfied. The red region corresponds to BBN excluded region, gray to overproduction of 
dark matter and brown to $\Delta N_{\rm eff}>1$. Red and brown solid lines show the boundaries of excluded 
regions due to BBN and dark radiation respectively.

\subsection{SUSY DFSZ model}

\begin{figure}[H]
\begin{center}
\includegraphics[height=0.29\textheight]{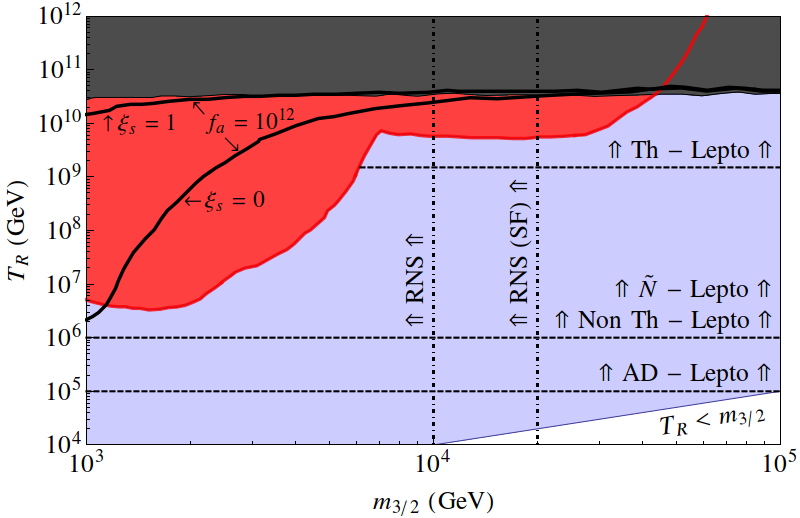}\\
\includegraphics[height=0.29\textheight]{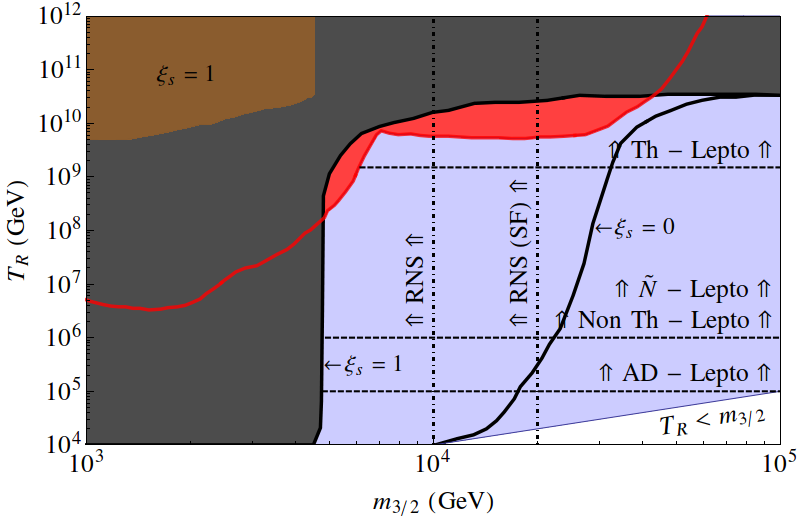}\\
\includegraphics[height=0.29\textheight]{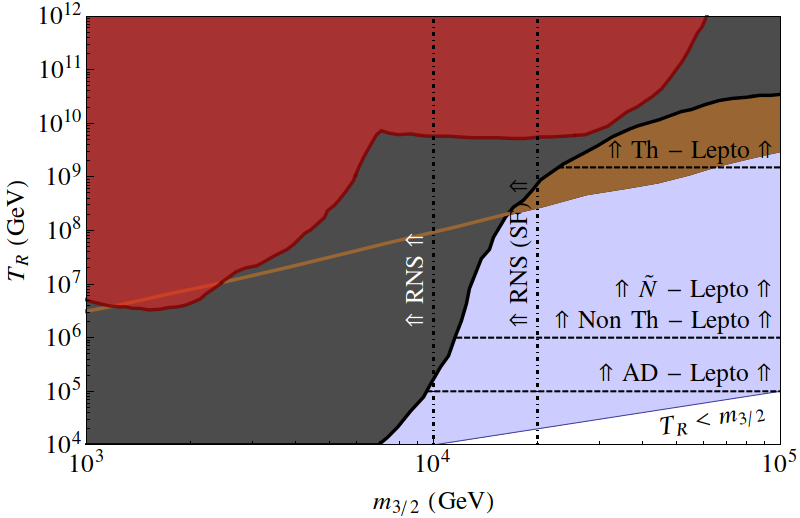}
\caption{
Plot of allowed regions in $T_R$ vs. $m_{3/2}$ plane in the SUSY DFSZ axion 
model for {\it a}) $f_a=10^{11}$ and $10^{12}$ GeV, 
{\it b}) $f_a=10^{13}$ GeV, for $\xi_s =0$ and 1 and {\it c}) $f_a=10^{14}$ GeV
for $\xi_s =1$. For $f_a=10^{11}$, $T_{R}>10^{11}$ is forbidden to avoid PQ symmetry restoration.
We take $m_s=m_{\ta}\equiv m_{3/2}$ in all plots.
\label{fig:dfsz}}
\end{center}
\end{figure}

Our first results of allowed regions in the $T_R$ vs $m_{3/2}$ plane are 
shown in Fig. \ref{fig:dfsz}. In frame {\it a}), we first take $f_a=10^{11}$ GeV and 
$10^{12}$ GeV and show allowed and excluded regions. 
For lower values of $f_a$, DM density is enhanced by gravitino decay only and BBN constraints are violated by 
late-decaying gravitinos since axinos and saxions are short-lived. For $f_a<10^{11}$ GeV, BBN bounds and DM exclusion contours can be read from Fig. \ref{fig:dfsz} once the region $T_R>f_a$ is omitted.
As we increase $f_a$ to $10^{11}$ GeV, then the axino and saxion decay rates are suppressed and they decay later. However, they still typically decay before neutralino freeze-out and thus do not change the picture. 

The gray band at the top of frame {\it a}) is forbidden due to overproduction of
WIMP dark matter due to thermal gravitino production and decay well after WIMP freeze-out.
This occurs for $T_R\agt 3\times 10^{10}$ GeV when $f_a=10^{11}$.
The red-shaded region occurs due to violation
of BBN constraints on late-decaying neutral relics. In the case of frame {\it a}), this comes again from gravitino production along with decay after the onset of BBN. Here, we use a digitized version of BBN 
constraints from Jedamzik~\cite{jedamzik} which appear in the $\Omega_Xh^2$ vs. $\tau_X$ plane 
where $X$ stands for the quasi-stable neutral particle, $\Omega_Xh^2$ is its would-be relic abundance 
had it not decayed and $\tau_X$ is its lifetime. The curves also depend on the $X$-particle  hadronic
branching fraction $B_h$ and on the mass $m_X$. Ref.~\cite{jedamzik} presents results for
$m_X=0.1$ and $1$ TeV and we extrapolate between and beyond 
these values for alternative mass cases. Together, the red- and gray-shaded regions constitute the
well-known {\it gravitino problem}: thermal gravitino production, which is proportional
to $T_R$~\cite{gravprod}, can lead to overproduction of decay-produced WIMPs or violations of BBN 
constraints.

In comparison, we also show several lines. The black vertical lines show the constraint
from naturalness ($\Delta_{\rm EW}<30$) that arises on $m_0$ in the NUHM2 model (labelled ``RNS'' at $m_{3/2}=10$ TeV for universal generations and ``RNS SF'' at $m_{3/2}=20$ TeV for split families~\cite{rns}). Here, we 
assume $m_{\tq}\simeq m_{3/2}$ where the gravitino mass sets the matter scalar mass scale most directly
in gravity-mediated SUSY breaking models.\footnote{In the numerical scans, 
the SUSY spectrum is fixed as we discussed in the beginning of this section. 
However, even if the scalar masses differ from the fixed benchmark values as the gravitino varies, 
dark matter and baryogenesis do not change significantly. 
Thus, this argument remains correct for this reason. }
In addition, we show the regions where
various leptogenesis mechanisms can account for the BAU. The region above $T_R=1.5\times 10^9$ GeV
is where thermal leptogenesis (THL) can occur. From the plot, we see the viable region,
colored as light-blue, is bounded by $m_{3/2}\agt 5$ TeV by BBN, by $m_{3/2}\alt 10$ TeV by
naturalness and by $1.5\times 10^9$ GeV$<T_R<5\times 10^9$ GeV by BBN and by successful baryogenesis.
Thus, THL is viable only in a highly restricted region of parameter space.
In contrast, non-thermal leptogenesis (NTHL) and sneutrino leptogenesis (OSL) are 
viable in a much larger region bounded from
below by $T_R\agt 10^6$ GeV while Affleck-Dine $LH_u$ flat-direction leptogenesis (ADL) is viable
in an even larger region for $T_R\agt 10^5$ GeV. These latter three leptogenesis regions are fully viable
for $m_{3/2}>1$ TeV. 

As $f_a$ is increased to $10^{12}$ GeV, then decays of axino and saxion are suppressed even further.
In this case, the DM-excluded region expands to the black contours labelled by $f_a=10^{12}$ GeV and $\xi_s =0$
or 1. The $\xi_s =1$ region is smaller than the $\xi_s =0$ region because for $\xi_s =1$ the saxion
decay width increases due to $s\to aa$ and the saxion lifetime is quicker.
The important point is that SUSY electroweak naturalness
expects $f_a\sim\sqrt{\mu M_P/\lambda_\mu}\sim 10^{10}-10^{12}$ GeV and for these values then there are wide
swaths of parameter space which support NTHL, OSL and ADL, and even THL is viable in some small region.

Instead, if we increase $f_a$ to $\sim 10^{13}$ GeV as in frame {\it b}), then we are somewhat beyond
the natural value of $f_a$, but also now the DM-forbidden region has increased greatly 
so that only values of $m_{3/2}\agt 5$ TeV are allowed for $\xi_s=1$, while for $\xi_s=0$ then
{\it all} of natural SUSY parameter space is forbidden. For low values of $m_{3/2}$($=m_{s}\Rightarrow$ long-lived saxions) and at high $T_{R}$, 
the decay $s\to aa$ produces too much dark radiation for $\xi_s=1$ case only. This region is colored brown and
triply excluded by DM, BBN and dark radiation constraints. In frame {\it c}), with
$f_a=10^{14}$ GeV, then natural SUSY parameter space is mostly forbidden by overproduction of
WIMPs for $\xi_s =1$ and totally forbidden for $\xi_s =0$ (not shown in the Fig. \ref{fig:dfsz}c).
In addition, the brown-shaded region ($\Delta N_{\rm eff}>1$) has extended and imposes an additional excluded region
for $m_{3/2}\gtrsim15$ TeV and $T_R\gtrsim10^8$ GeV.

These results have important implications for axion detection. Currently, the ADMX experiment is
exploring regions of $f_a/N_{\rm DW}\agt 10^{12}$ GeV ($N_{\rm DW}$ is domain wall number). 
Future plans include an exploration of regions down to
$f_a/N_{\rm DW}\agt 10^{11}$ GeV. To make a complete exploration of the expected locus of the axion
in natural SUSY, then such experiments should also aim for exploration down to $f_a/N_{\rm DW}\sim 10^{10}$ GeV.
For even smaller $f_a/N_{\rm DW}<10^{10}$ GeV values, then axion CO-production requires $\theta_i$ values very close 
to $\pi$ and the axion production rates would be considered as fine-tuned~\cite{bbbss}.

\subsection{SUSY KSVZ model}
In this subsection, we show baryogenesis-allowed regions in the $T_R$ vs. $m_{3/2}$ plane for 
the SUSY KSVZ model. We regard the SUSY KSVZ model as less lucrative in that one loses
the DFSZ solution to the SUSY $\mu$ problem and the connection with electroweak 
naturalness. In addition, if the exotic heavy quark field $Q$ is not an element of a
complete GUT multiplet, then one loses gauge coupling unification. 
Nonetheless, it is instructive to view these results for comparison to the SUSY DFSZ case.
\begin{figure}[H]
\begin{center}
\includegraphics[height=0.29\textheight]{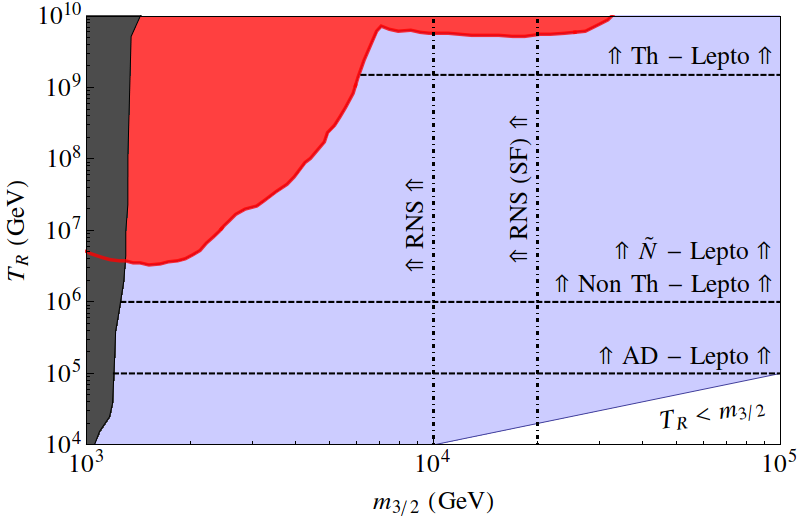}\\
\includegraphics[height=0.29\textheight]{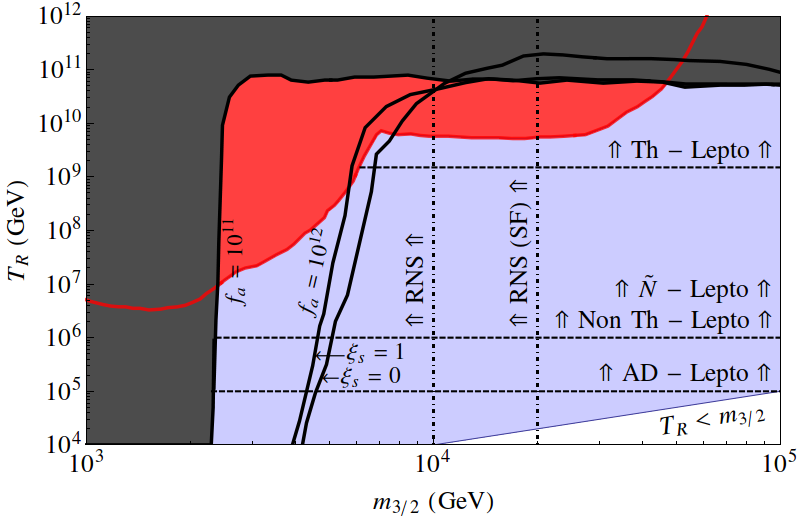}\\
\includegraphics[height=0.29\textheight]{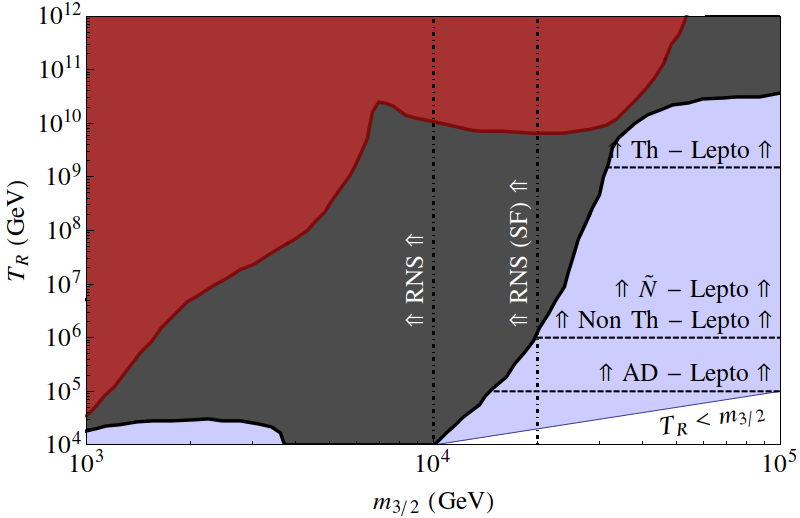}
\caption{
Plot of allowed regions in $T_R$ vs. $m_{3/2}$ plane in the SUSY KSVZ axion 
model for {\it a}) $f_a=10^{10}$ GeV, {\it b}) $10^{11}$ and $10^{12}$ GeV for $\xi_s =$0 and 1 and {\it c}) 
$f_a=10^{13}$ GeV for $\xi_s =0$.  For $f_a=10^{11}$, $T_{R}>10^{11}$ is forbidden to avoid PQ symmetry restoration. We take $m_s=m_{\ta} =m_{3/2}$ in all plots.
\label{fig:ksvz}}
\end{center}
\end{figure}

In Fig. \ref{fig:ksvz}{\it a}), we show results for $f_a=10^{10}$ GeV.
Even for $f_a$ as low as $10^{10}$ GeV, the gray-shaded WIMP-overproduction region
occupies the region with $m_{3/2}\alt1.3$ TeV.
In this region, since $m_{\ta}\simeq m_{3/2}$, then thermal axino production followed by decay after
neutralino freeze-out leads to WIMP over production across a wide range of $T_R$ values.
This is because the axino decay is suppressed by $Q$-mediated loops as compared to SUSY DFSZ.
As $f_a$ is increased to $10^{11}$ GeV (Fig. \ref{fig:ksvz}{\it b})), then the DM-forbidden region expands out to 
$m_{3/2}\sim 2$ TeV region. For $f_a=10^{12}$ GeV (Fig. \ref{fig:ksvz}{\it b})), then the DM-forbidden region expands out to
$m_{3/2}\sim 4$ TeV. Even for this high value of $f_a$, there is still room for 
leptogenesis in natural SUSY models for each of the cases of THL, NTHL, OSL and ADL.\footnote{For this case only, 
we have found that there exists some mild entropy dilution $r$ of 
$n_L$ due to thermal axino production for 
$T_R\sim 10^{10}-10^{11}$ GeV by up to a factor of 2.
Since these $T_R$ values are beyond the lower limit, our plots hardly change.
Alternatively, the THL lower bound on $T_R$ may be interpretted as a lower 
bound on $T_R/r$.}

For the SUSY KSVZ model with $f_a=10^{13}$ GeV as shown in Fig. \ref{fig:ksvz}{\it c}), 
then the DM-forbidden region has expanded to exclude all viable natural SUSY parameter space
except for a tiny slice with $m_{3/2}\sim 15-20$ TeV and $T_R<10^6$ GeV where ADL might still function.

\section{Constraints in the $T_R$ vs. $f_a$ plane for fixed $m_{3/2}$}
\label{sec:tr_fa}

In this section, we examine the DM constraints on baryogenesis in the $T_R$ vs. $f_a$ plane
for fixed natural $m_{3/2}$ values to gain further insights on axion decay constant dependence of 
the constraints for $T_{R}$ between $10^{4}$ - $10^{12}$ GeV. 
On these planes, 
in the yellow region labelled $T_R>f_a$ we expect PQ symmetry to be restored during reheating which leads to
generation of separate domains with different $\theta$ values and the appearance of
domain walls and associated problems. In this case, axion coherent oscillations must be averaged over separate domains~\cite{axdm}. As before, we do not consider this region.

\subsection{SUSY DFSZ model}

\begin{figure}[tbp]
\begin{center}
\includegraphics[height=0.30\textheight]{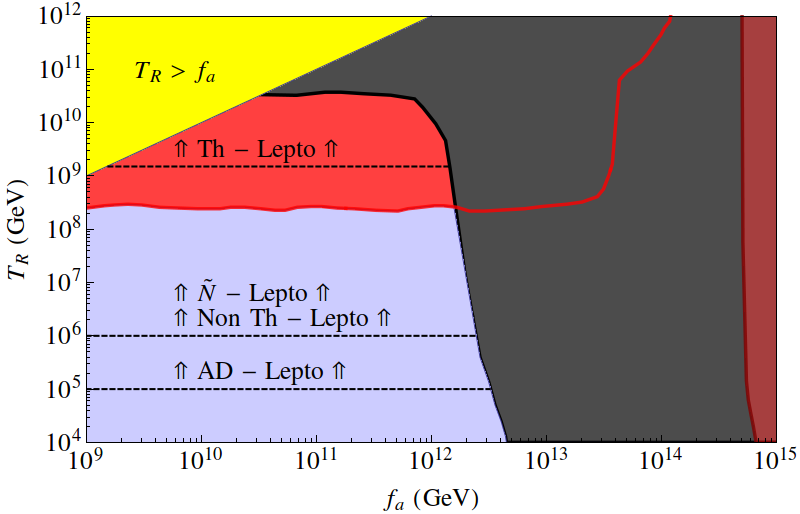}
\includegraphics[height=0.30\textheight]{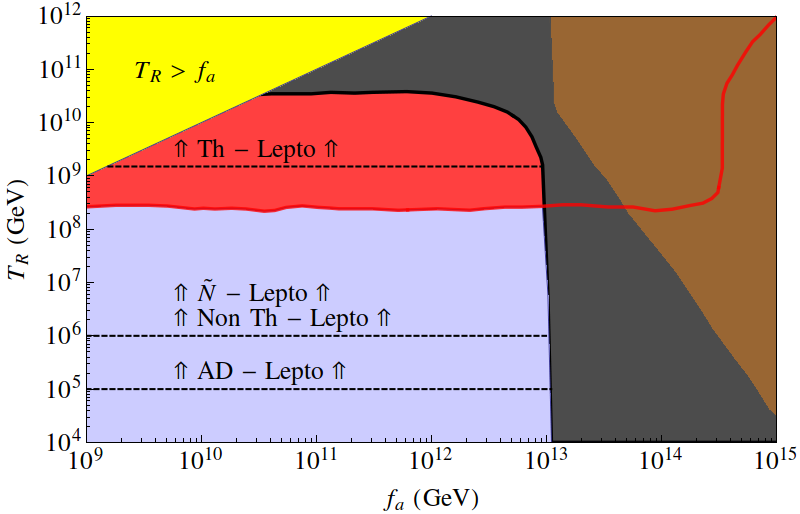}
\caption{
Plot of allowed regions in $T_R$ vs. $f_a$ plane in the SUSY DFSZ axion 
model for $m_{3/2}=5$ TeV and with {\it a}) $\xi_s =0$ and {\it b}) $\xi_s =1$.
We take $m_s=m_{\ta}= m_{3/2}$ in all plots.
\label{fig:d_fa_5}}
\end{center}
\end{figure}

In Fig. \ref{fig:d_fa_5}, we plot allowed and forbidden regions for baryogenesis
in SUSY DFSZ model in the $T_R$ vs. $f_a$ plane for $m_{3/2}=5$ TeV. 
In frame {\it a}), with $\xi_s =0$, the gray-shaded region still corresponds to WIMP 
overproduction and sets an upper limit of $f_a\alt 10^{12}$ GeV. The red-shaded region 
corresponds to violation of BBN constraints from
late decaying gravitinos and bounds $T_R$ from above: $T_R\alt 2\times 10^8$ GeV which
excludes the possibility of THL. Still, large regions of natural SUSY parameter space
are consistent with NTHL, OSL and with ADL. 
The BBN bound kicks in again at $f_a\sim6\times10^{14}$ due to long-lived saxions.
For the case of $\xi_s =1$ shown in Fig. \ref{fig:d_fa_5}{\it b}), then $s\to aa$ is turned on. 
This leads to the brown dark-radiation excluded region at very large $f_a$ values and large $T_R$.
In addition, we note for this case that the red-shaded BBN forbidden region has actually expanded
compared to frame {\it a}).
This is because for $\xi_s =0$, the CO-produced saxions inject considerable entropy into the cosmic
soup at large $f_a$ thus diluting the gravitino abundance. For $\xi_s =1$, then the saxion decays
more quickly leading to less entropy dilution of gravitinos and thus more restrictive BBN bounds.
Thus, the BBN constraints are actually more severe for $\xi_s =1$.
In addition, for frame {\it b}), we see WIMP overproduction bounds are less severe with
$f_a\alt 10^{13}$ GeV being required for the allowed regions. These are due to a reduced
$s\to SUSY$ branching fractions for the $\xi_s =1$ case.

\begin{figure}[tbp]
\begin{center}
\includegraphics[height=0.30\textheight]{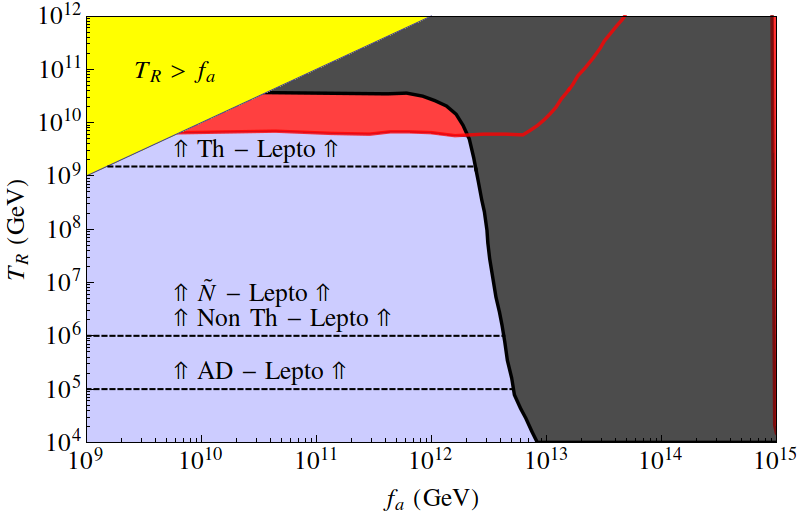}
\includegraphics[height=0.30\textheight]{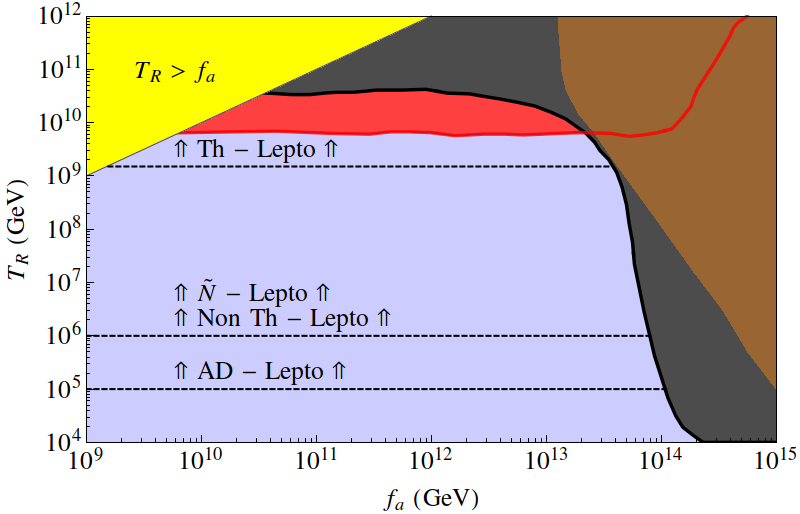}
\caption{
Plot of allowed regions in $T_R$ vs. $f_a$ plane in the SUSY DFSZ axion 
model for $m_{3/2}=10$ TeV and with {\it a}) $\xi_s =0$ and {\it b}) $\xi_s =1$.
\label{fig:d_fa_10}}
\end{center}
\end{figure}

In Fig. \ref{fig:d_fa_10} we show allowed and excluded regions in the $T_R$ vs. $f_a$ plane for
$m_{3/2}=10$ TeV. In the case of $\xi_s =0$ shown in frame {\it a}), 
the larger gravitino mass causes the gravitinos to decay more quickly 
so that BBN constraints are diminished: in this case, the THL scenario with $T_R>1.5\times 10^9$ GeV
is allowed in contrast to the previous case with $m_{3/2}=5$ TeV. In addition, broad swaths
of parameter space are allowed for the NTHL, OSL and ADL scenarios with $f_a\alt 5\times 10^{12}$ GeV.
For larger $f_a$ values, then axino and saxion production followed by late decays leads to too much WIMP 
dark matter.
For the case with $\xi_s =1$ shown in frame {\it b}), we see again the BBN constraints are
somewhat enhanced
due to diminished entropy dilution of gravitinos at large $f_a$.
In addition, a dark-radiation forbidden region has appeared. Most importantly, the DM-allowed region occurs
for $f_a\alt 10^{14}$ GeV so that large swaths of parameter space are open for baryogenesis. 
This is because, since we take $m_{\ta}=m_s=m_{3/2}$, then the axinos and saxions are also shorter-lived
and tend to decay earlier - frequently before WIMP freeze-out - so DM overproduction is more easily avoided.

For even larger values of $m_{3/2}$ up to $m_{3/2}\sim25$ TeV, we would expect to see a very similar BBN 
constraint since BBN bounds are not sensitive to any changes in $m_{3/2}$ for $7\mbox{ TeV}\alt m_{3/2}\alt 25$ TeV 
(see Fig. \ref{fig:dfsz}). As $m_{3/2}$ increases and reaches beyond $m_{3/2}\sim65$ TeV, then gravitino decays much sooner and does not violate BBN constraints at all. However DM production highly depends on $f_a$ and the DM exclusion picture would look different up to a maximum $f_a$ after which the whole parameter space is excluded by 
too much DM.
\subsection{SUSY KSVZ model}
\begin{figure}[tbp]
\begin{center}
\includegraphics[height=0.30\textheight]{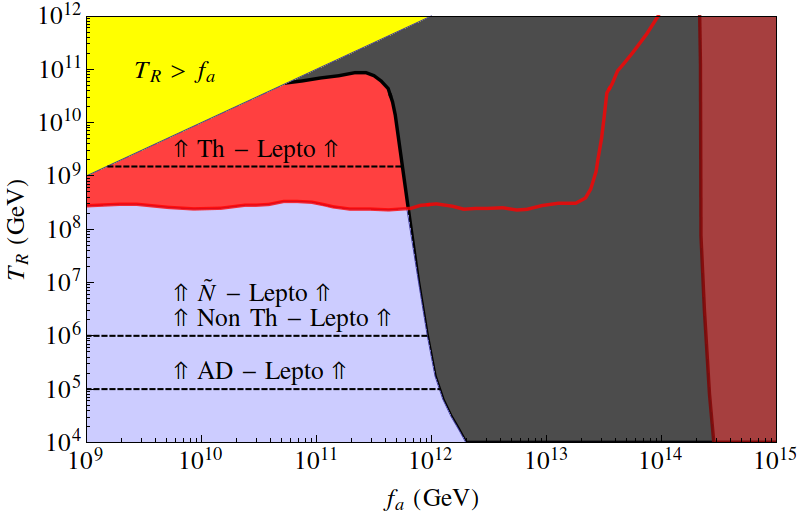}
\includegraphics[height=0.30\textheight]{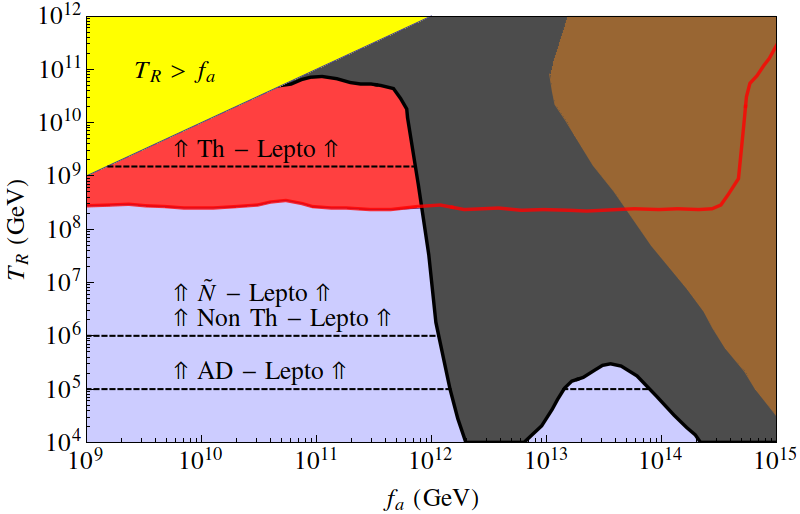}
\caption{
Plot of allowed regions in $T_R$ vs. $f_a$ plane in the SUSY KSVZ axion 
model for $m_{3/2}=5$ TeV and with {\it a}) $\xi_s =0$ and {\it b}) $\xi_s =1$.
\label{fig:k_fa_5}}
\end{center}
\end{figure}
In this subsection, we show corresponding results in the $T_R$ vs. $f_a$ plane for SUSY KSVZ.
In Fig. \ref{fig:k_fa_5} we show the plane for $m_{3/2}=5$ TeV and {\it a}) $\xi_s =0$.
Here, we see that THL is ruled out due to the severe BBN bounds arising from gravitino
production and decay which restrict $T_R\alt 2\times 10^8$ GeV while the DM restriction 
rules out $f_a\agt 10^{12}$ GeV. The NTHL, OSL and ADL are still viable baryogenesis mechanisms 
over a wide range of $T_R$ and $f_a$ values.
In frame \ref{fig:k_fa_5}{\it b}) for $\xi_s =1$, the DM forbidden region is similar with
a $f_a<10^{12}$ GeV restriction. However, the BBN restricted region has increased because
there is less entropy dilution from saxion decay of the gravitinos abundance. The expanded BBN region 
lies in the already DM and dark radiation excluded region so provides no additional constraint.
Since saxions decay earlier for $\xi_s =1$ compared to $\xi_s =0$, then they inject neutralinos at
a higher decay temperature $T_s^D$; as a consequence, a small DM-allowed region appears at high $f_a\sim 10^{13}-10^{14}$ GeV
and $T_{R}\sim 10^5$ GeV which is barely consistent with ADL.
\begin{figure}[tbp]
\begin{center}
\includegraphics[height=0.30\textheight]{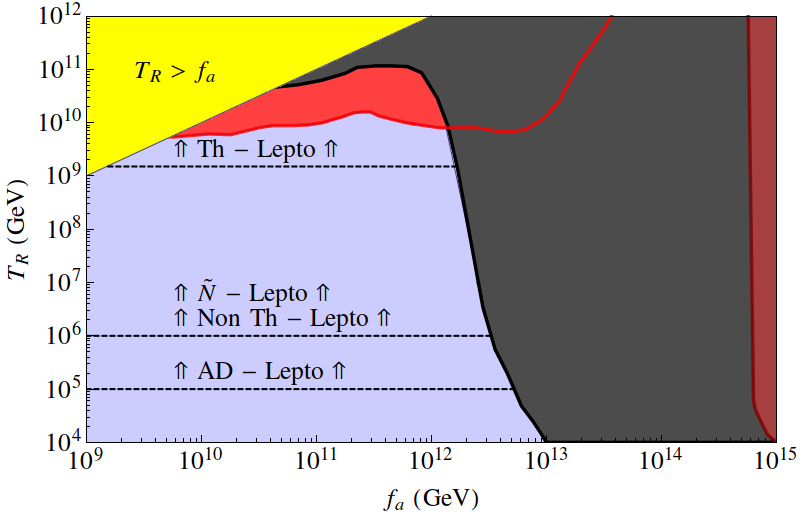}
\includegraphics[height=0.30\textheight]{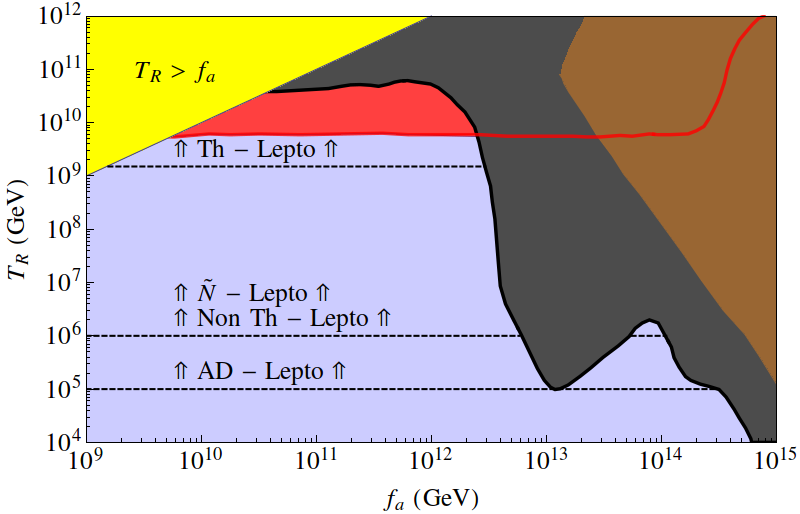}
\caption{
Plot of allowed regions in $T_R$ vs. $f_a$ plane in the SUSY KSVZ axion 
model for $m_{3/2}=10$ TeV and with {\it a}) $\xi_s =0$ and {\it b}) $\xi_s =1$.
\label{fig:k_fa_10}}
\end{center}
\end{figure}
In Fig. \ref{fig:k_fa_10}{\it a}), we show the same $T_R$ vs. $f_a$ plane with $\xi_s =0$, but this time for
a heavier value of $m_{3/2}=m_s=m_{\ta}=10$ TeV. The higher value of $m_{3/2}$ means gravitinos decay 
more quickly and at higher temperature so that the BBN bound on $T_R$ is given by $T_R\agt4\times 10^9$ so
that THL is again viable. Also, the DM-allowed region has moved to a higher $f_a$ bound of $f_a\alt 2\times 10^{12}$ GeV.
In this frame, all four baryogenesis mechanisms are possible. 
In Fig. \ref{fig:k_fa_10}{\it b}), we show the same plane for $\xi_s =1$. Here a prominent dark 
radiation excluded region appears at large $f_a\agt 10^{13}-10^{14}$ GeV, although this region is already
excluded by WIMP overproduction and by BBN. The larger saxion width arising from the additional $s\to aa$ decay mode
means the saxion decay at higher temperatures leading to some possible allowed regions appearing at $f_a\sim 10^{14}$ GeV
and $T_R\sim 10^5$ GeV which admits ADL. Otherwise, large regions of viable parameter space exists for
$f_a\alt 2\times 10^{12}$ GeV and for $T_R\alt 4\times 10^9$ GeV where all four leptogenesis mechanisms are possible.

\section{Conclusion}
\label{sec:conclude}

In this paper we have investigated constraints on four compelling baryogenesis-via-leptogenesis
scenarios within the framework of supersymmetric models with radiatively-driven naturalness.
These models are especially attractive since they contain solutions to the gauge hierarchy problem (via SUSY), 
the strong CP problem (via the axion), the SUSY $\mu$ problem (for the case of the SUSY DFSZ axion) 
and the Little Hierarchy problem (where $\mu\sim 100-200$ GeV is generated from multi-TeV values of $m_{3/2}$).
The characteristic, unambiguous signature of such models is the presence of light higgsinos $\tz_{1,2}$ 
and $\tw_1^\pm$ with mass $\sim \mu$. In these models, the LSP is a higgsino-like WIMP 
which is thermally underproduced. The remainder of the dark matter abundance is filled by the axion.
Indeed, over most of parameter space the axion forms the bulk of dark matter~\cite{bbc}.

In supersymmetric dark matter models, then baryogenesis mechanisms are confronted by the gravitino
problem: gravitinos which are thermally produced in the early universe can lead to overproduction of WIMPs
or to violations of BBN constraints. In SUSY axion models, there are analogous problems arising from
thermal axino production and decay and from thermal- and oscillation-produced saxions.
We calculated regions of the $T_R$ vs. $m_{3/2}$ plane in the compelling RNS SUSY model 
with DFSZ axions and $\xi_s =0$ and 1. Our main result is that the region of parameter space preferred
by naturalness with $f_a\sim\sqrt{\mu M_P/\lambda_\mu}\sim 10^{10}-10^{12}$ GeV supports all four leptogenesis
mechanisms. The thermal leptogenesis is perhaps less plausible since its allowed region is nestled
typically between the constricted region of 
$7\mbox{ TeV}<m_{3/2}<10$ TeV (or $<20$ TeV in RNS for split families) and 
$1.5\times 10^9\ {\rm GeV}<T_R<4\times 10^9$ GeV. The other NTHL, OSL and ADL mechanisms
can freely operate over a broad region of parameter space for $f_a\alt 10^{12}$ GeV and $T_R\agt 10^5$ GeV.
We also evaluated all constraints in the $T_R$ vs. $f_a$ plane for fixed $m_{3/2}=5$ and 10 TeV.

The broad allowed regions of parameter space basically favor the following. 
\bi
\item  Multi-TeV values of $m_{3/2}$ to avoid BBN constraints and to hasten saxion and axino decays.
Since $m_{3/2}$ sets the scale for superpartner masses at LHC, these multi-TeV values of $m_{3/2}$ are also supported
by LHC8 sparticle search constraints and the large value of $m_h\sim 125$ GeV at little cost to naturalness.
\item A value of $f_a \sim 10^{10}-10^{12}$ GeV which suppresses WIMP over production from axino/saxion production.
Such values of $f_a$ lead to axion masses somewhat above the standard search region of ADMX and should motivate future
axion search experiments to increase their search region to heavier axion masses.
\item Values of $T_R\sim 10^5-10^9$ GeV.
\ei
For completeness, we have also evaluated the leptogenesis allowed regions in the SUSY KSVZ model for 
which an alternative solution to the $\mu$ problem is needed. The loop-suppressed axino and saxion decay rates
typically lead to more stringent constraints in this case although regions of parameter space can still be
found where the various leptogenesis mechanisms are still possible.

\acknowledgments

We thank Vernon Barger for discussions and Andre Lessa for earlier collaborations on these topics.
This work was supported in part by the US Department of Energy, 
Office of High Energy Physics.
The computing for this project was performed at the 
OU Supercomputing Center for Education \& Research (OSCER) at the University of
Oklahoma (OU). 
KJB is also supported by Grant-in-Aid for Scientific research No. 26104009.

\end{document}